\documentclass{spie}

\usepackage{amsmath,amsfonts,amssymb}
\usepackage{graphicx}
\usepackage[colorlinks=true, allcolors=blue]{hyperref}
\usepackage{sidecap}
\usepackage[dvipsnames]{xcolor}
\usepackage{multirow}

\def\deg{\hbox{$^\circ$}}

\def\lae{\mathrel{\raise .4ex\hbox{\rlap{$<$}\lower 1.2ex\hbox{$\sim$}}}}
\def\gae{\mathrel{\raise .4ex\hbox{\rlap{$>$}\lower 1.2ex\hbox{$\sim$}}}}

\newcommand{\rs}{{\em REDSoX}}

\title{Design and testing progress towards the first flight of the rocket experiment demonstration of a Soft X-ray Polarimeter (REDSoX)}

\author[a]{Sarah N.\ T.\ Heine}
\author[a]{Herman L.\ Marshall}
\author[a]{Alan Garner}
\author[a] {F. Elio Angile}
\author[a]{C. Marc Baker}
\author[b] {Stephen Bongiorno}
\author[a]{Leonardo Drake}
\author[a] {H. Moritz G\"unther}
\author[a] {Ralf K.~Heilmann}
\author[a] {Beverly LaMarr}
\author[a] {Sean McNeil}
\author[a] {Jill Juneau}
\author[a] {Swati Ravi}
\author[c] {Eric Gullikson}

\affil[a]{MIT Kavli Institute, Cambridge, MA, USA}
\affil[b] {NASA Marshall Space Flight Center, Huntsville, AL, USA}
\affil[c] {Lawrence Berkeley National Laboratory, Berkeley, CA, USA }

\authorinfo{Further author information: (Send correspondence to S.N.T.H.)\\S.N.T.H..: saraht@mit.edu,
Telephone: 1 617 258 8819} 

\begin{document}

\maketitle

\begin{abstract}
The Rocket Experiment Demonstration of a Soft X-ray Polarimeter (REDSoX\footnote{The name of the instrument
is used with the permission of the Red Sox Baseball Club and Major League Baseball.}) is a NASA-funded sounding rocket mission.  The rocket payload will measure polarization strength and direction as a function of energy in the 0.2-0.4 keV band, providing complementary measurements to those made by IXPE in the 2-8 keV band.  The first flight, scheduled for late 2027, will provide a technology demonstration of our polarimeter concept, which utilizes an aligned system of a focusing optic, Critical-Angle Transmission (CAT) gratings, Laterally Graded Multilayer (LGML) mirrors, and Charge Coupled Device (CCD) detectors to measure polarization.   We will describe the design of the instrument post-critical design review, the status of flight hardware testing, and payload assembly.
\end{abstract}


\section{REDSoX Mission Overview}
The Rocket Experiment Demonstration of a Soft X-ray (REDSoX) polarimeter is a NASA-funded sounding rocket mission planned for launch in late 2027 \cite{redsoxjatis,redsox24,REDSoXSPIE2025}.  REDSoX will measure the linear polarization of X-ray light between 200 and 400 eV with a minimum detectable polarization of roughly 20$\%$.  The first planned target is MK421, which is chosen in large part because of its brightness.  By measuring polarization in the 200-400 eV band, REDSoX will complement similar measurements made by the Imaging X-ray Polarimetry Explorer (IXPE) in the 2-8 keV band.  A key future application for this technology is the study of isolated neutron stars and their magnetized atmospheres, which mainly emit below 1 keV.  This science is inaccessible from a sounding rocket mission because of the short exposure times available; however, this science will be well within reach of GOSoX, an orbital mission recently selected as a NASA Pioneer utilizing the same technology \cite{GOSoXSPIE2026}.

The REDSoX mission team completed a critical design review in January 2026, and is currently working to test major components and finalize design details prior to assembly of the payload.  The flight is scheduled for late 2027.  Here we provide an update on the status of the REDSoX mission development.  

\subsection{Working Principle and Design Overview}

\begin{figure}
   \centering
    \includegraphics[width=13cm]{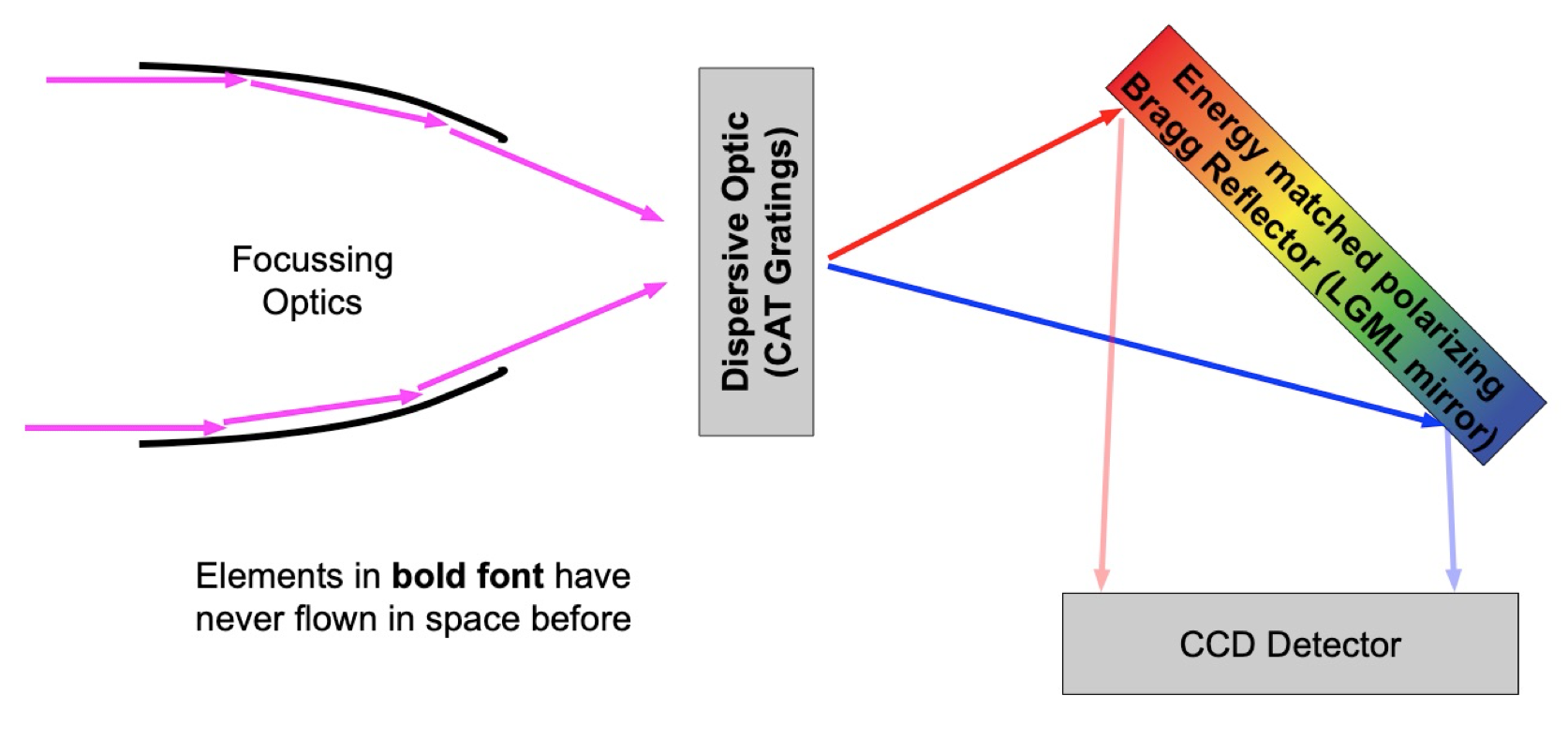}
 \caption{A diagram of the main optical elements of a single (one of three) channel of the REDSoX payload, illustrating how a spectral polarization measurement is achieved.  No optical element comparable to the CAT gratings or the LGML mirrors has been flown in space, so this mission will serve to raise technology readiness levels in addition to its science, training, and technology development goals.}
\label{fig:workingprincipal}
\end{figure}

REDSoX provides spectropolarimetry measurements across our bandpass by combining precisely aligned optical elements.  Incoming light is incident on the 5 shells of replicated Ni-shell Wolter-I X-ray mirrors, which focus the light with a focal length of 2.5 meters.  There are three polarimetry channels, each of which utilizes two petals (an upper and a lower) of critical-angle transmission (CAT) gratings placed within the converging beam to disperse the light onto a laterally graded multilayer (LGML) mirror.  Our LGMLs have a layer spacing that is varied along the surface of the mirror such that the energy of the Bragg peak across the surface of the mirror varies linearly.  We align the LGMLs with the CAT gratings such that the energy of the light dispersed in first order by the two  CAT gratings petals matches the energy of the Bragg peak at each point along the surface of the LGML.  The 45 degree Bragg reflection off of the LGML reflects light polarized along the surface of the multilayer, which is then measured by a CCD.  Light that is not polarized along the surface of the mirror will not be reflected.  The three LGMLs are oriented at 120 degrees to each other, which will allow us to measure both the strength and the direction of polarization.  A diagram of a single channel is shown in Figure \ref{fig:workingprincipal}.  For REDSoX, the focusing optics will be produced by NASA Marshall Space Flight Center, the CAT gratings have been provided by Izentis LLC in collaboration with the MIT Space Nanotechnology Laboratory, and the LGMLs were produced by Lawrence Berkeley National Laboratory.  The CCD cameras are being provided by XCAM Ltd., utilizing Teledyne-e2V CCD sensors.

\begin{figure}
   \centering
    \includegraphics[width=17cm]{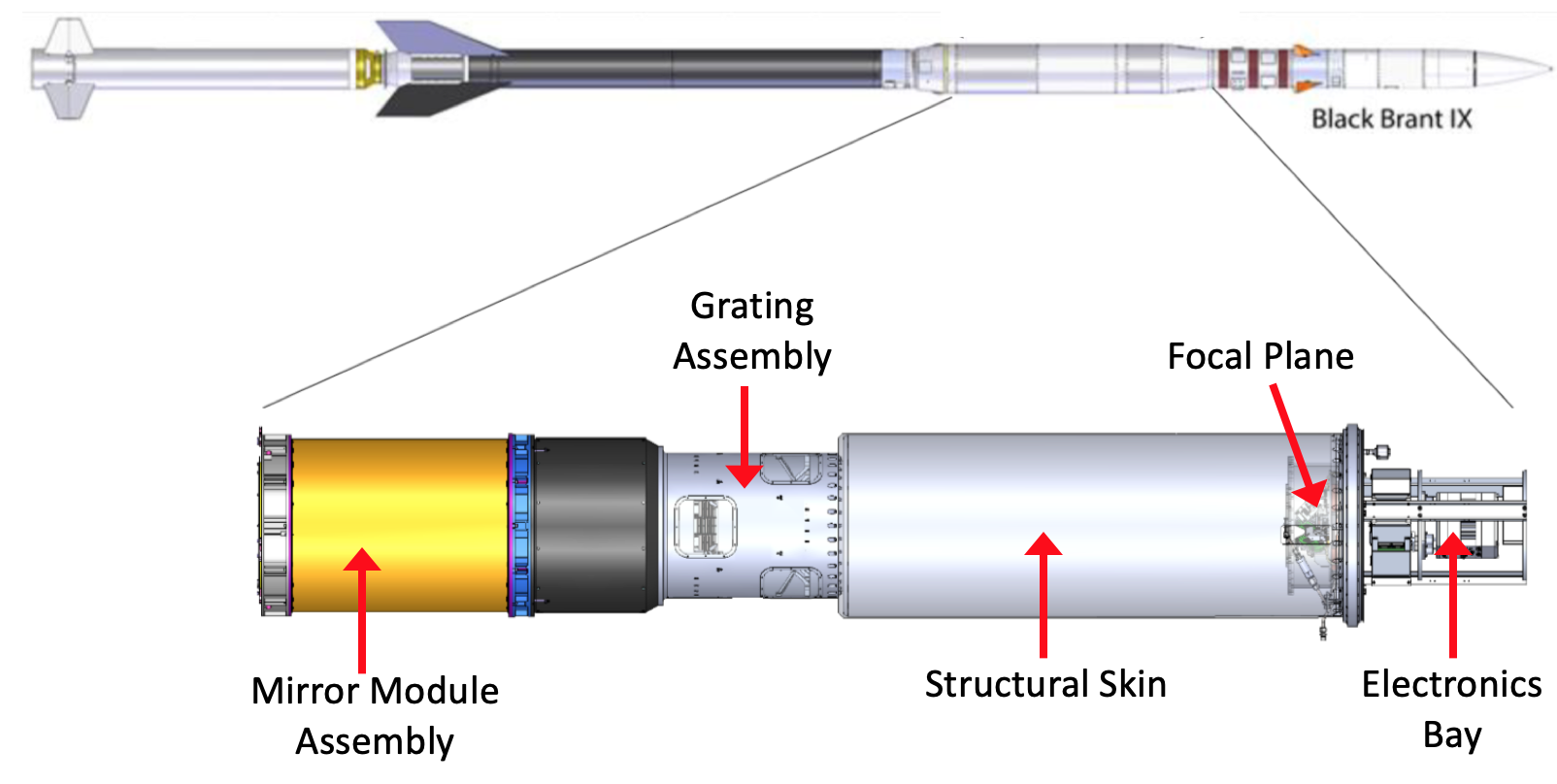}
 \caption{CAD representations of the full rocket (top) and the REDSoX payload (bottom).  Major systems and components are labeled.  The design is nearly finalized.}
\label{fig:REDSoXOverview}
\end{figure}

\begin{figure}
   \centering
    \includegraphics[width=16cm]{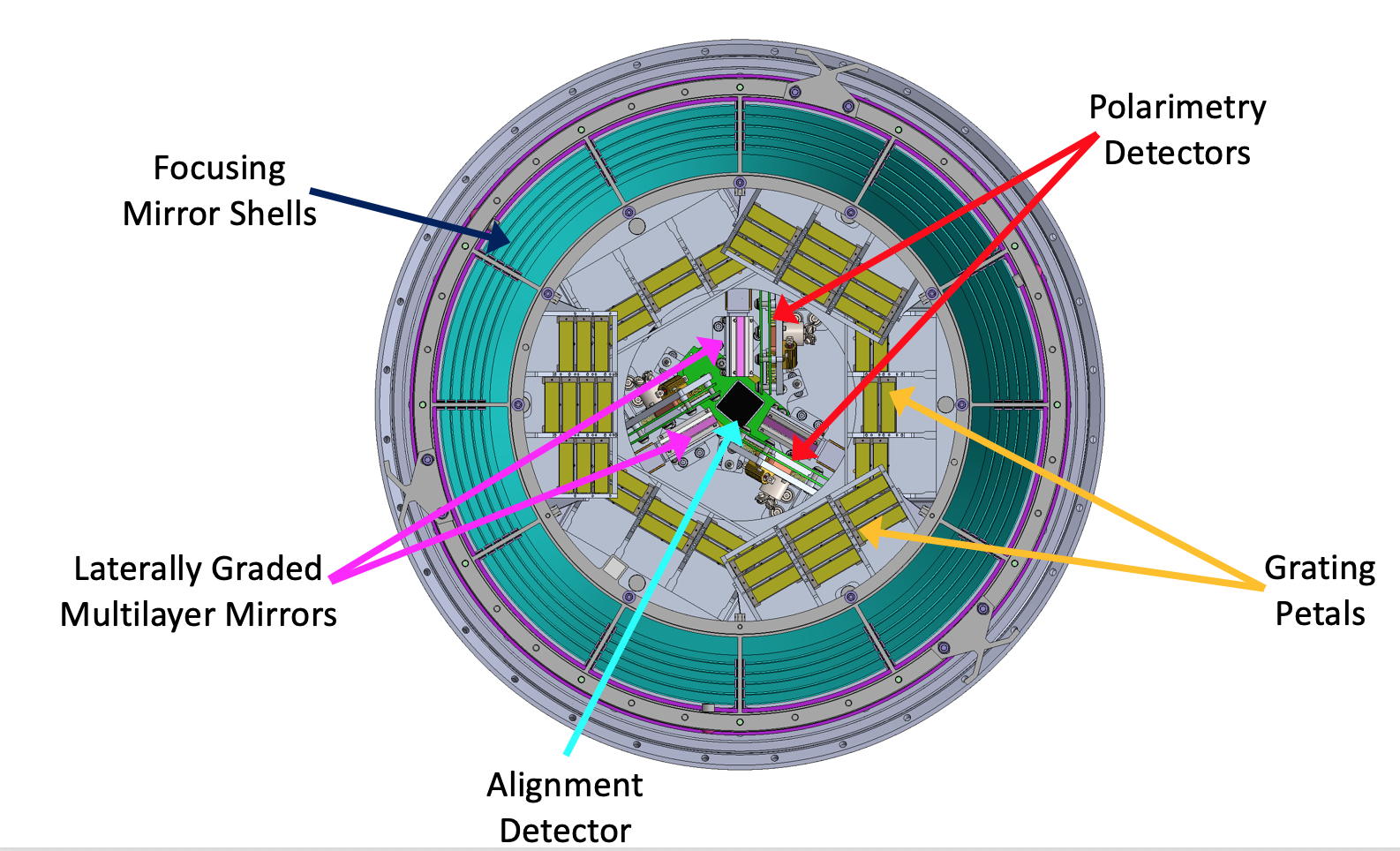}
 \caption{A CAD representation of the payload as seen from the aft end where photons enter the payload  The focusing mirror is shown in cyan, the gratings are shown in gold, the LGMLs are highlighted purple, and the detectors are black with their associated  headboards shown in green.  Major components are labeled.}
\label{fig:REDSoXEndview}
\end{figure}

CAD drawings of the rocket and the REDSoX payload are shown in Figures \ref{fig:REDSoXOverview} and \ref{fig:REDSoXEndview}.  These schematics show both a cutaway view of the payload emphasizing its structure, as well as an end view, which highlights the relative orientations of the optical elements of the payload.  In Figure \ref{fig:REDSoXOverview}, the focal plane is shown with the `top hat,' which is a cylindrical enclosure mounted on the room temperature bulkhead and enclosing the whole of the focal plane.  This acts as a light baffle, as well as a mount for the optical blocking filters, which will remain warm.  This will make them less susceptible to icing.  It also acts as a passive contamination shield for the cold multilayers and detectors on the focal plane.  In Figure \ref{fig:REDSoXEndview}, the top hat and optical blocking filters have been suppressed to allow better view of the optical components.   The star tracker, which will sit in the center of the focusing optic, has also been suppressed.  The focal plane will be held at the correct distance from the gratings module and the optics by a structural skin, which will be fabricated by NSROC and nested within the outer skin of the rocket.  

The central ``alignment detector'' is used to achieve sufficient pointing accuracy.  We require pointing precision to within 5 arcseconds.  This is achievable with the Wallops tri-level pointing system, which provides pointing to within $\pm$3 arcseconds at three sigma.  However, data suggests that the initial pointing is far less accurate due to factors including potential shifting of the alignment between the optical axis of the telescope and the star tracker during launch.  To correct for initial pointing errors we will spend the first 30 seconds of the flight acquiring an image of the source on the imaging detector.  The detector is defocused due to space constraints within the payload, but we will still be able to  use centroid data from this detector to guide a pointing adjustment to get the telescope within its pointing requirement for science acquisition during the remainder of the flight.  We also plan for a 60$\deg$ roll of the instrument halfway through the science acquisition time to allow us to remove systematic errors introduced by any efficiency differences in our channels from our final science result.

\section{MIT Polarimetry Beamline}

\begin{figure}
   \centering
    \includegraphics[width=17cm]{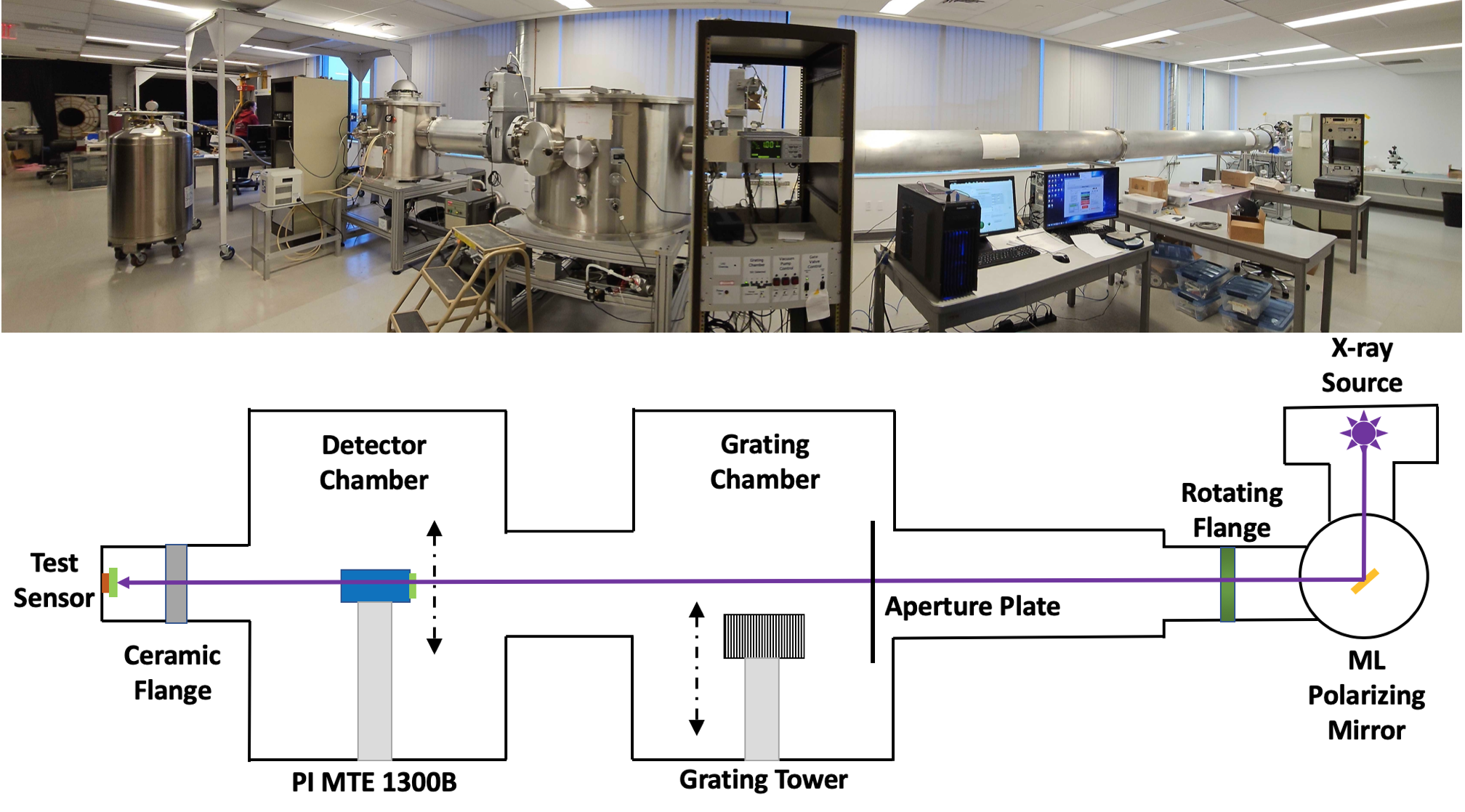}
 \caption{
\label{fig:Beamlinediagram}
A panoramic image of the polarimetry beamline (upper).  A block diagram of the MIT polarimetry beamline (lower).  The PI MTE 1300B detector is installed on a motorized X-Y stage and can be inserted into the beam or moved out.  It is installed with an optical blocking filter as well as a shutter, which closes during readout.  There is a an electrically isolating ceramic flange on the end of the beamline, on which a flange-mounted detector or small chamber with a mounted sensor chip may be connected for testing.
Gratings can be inserted and manipulated within the beam by motorized stages under vacuum and the PI detector can be moved to sample the dispersion of light off of the gratings.}
\end{figure}

Testing of most of the elements of REDSoX is being conducted in the MIT polarimetry beamline \cite{murphy:77322Y,heine17,garner19}.  This facility has been re-purposed from its original use as a calibration facility for the Chandra HETG gratings and now serves as a testbed for REDSoX hardware and other technology development projects.  A block diagram of the beamline is shown in Figure \ref{fig:Beamlinediagram}.  
   
 The beamline is roughly 20 meters long and has three chambers (source, grating and detector).  The X-ray source is a Manson source with several interchangeable anodes to produce various line energies.  The source light is reflected off of a W/B$_4$C laterally graded multilayer mirror installed on a motorized linear stage.  By selecting the anode material and the corresponding location to place the X-ray spot on the LGML (where the energy of the emission line produced by the source matches the energy of the Bragg peak on the LGML), we reflect polarized, monochromatic light at the energy of our choice down the beamline.  We are able to tune the mirror to several different emission lines between 183 eV and 700 eV without breaking vacuum.  The beamline can also produce energies between emission lines of the anodes by utilizing continuum emission.  This emission is not as strong as that of the emission lines, but has proven sufficient for testing given sufficient exposure time.  A rotating flange between the source assembly and the rest of the beamline allows us to rotate the source assembly via a motor to change the direction of polarization of the light traveling down the beamline while under vacuum.  The grating chamber contains an aperture plate, slit plate, and grating plate, which are all mounted on motorized stages.  Gratings installed on the grating plate with custom mounting handles can be inserted, manipulated, or removed from the beam using an additional motorized stage.

The detector chamber contains a Princeton Instruments MTE1300B in-vacuum CCD mounted on a motorized X-Y stage (perpendicular to the system optical axis).  This allows us to move the Princeton Instruments detector in and out of the beam while under vacuum.  Quick-look analysis is provided by custom Python-based code.  The motors and camera are controlled by custom Labview software.  There is also a flange on the end of the detector chamber allowing a test detector to be mounted in the beam path.  A detailed description of detector testing in the beamline can be found in Ref \citenum{DetectorSPIE2025}.

\vskip -0.1in

\section{Mirror Module Assembly}
\begin{figure}
    \centering
    \includegraphics[width=15 cm]{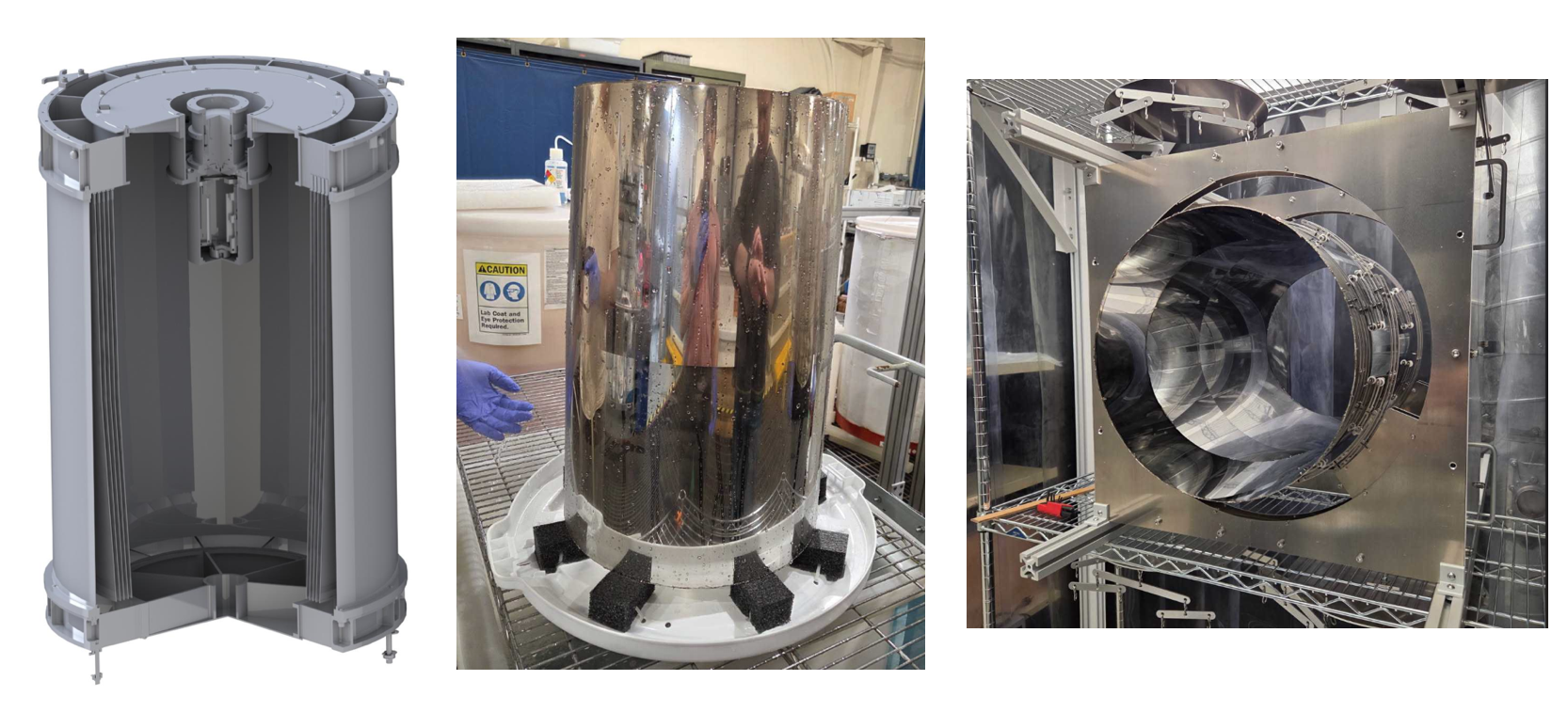}
 \caption{\textit{Left}: CAD rendering of the MMA, \textit{Center}: The first M5 shell replicated off of the REDSoX mandrels, \textit{Right}: The REDSoX M5 shell suspended in a mount for testing in the MSFC 100 meter beamline.}
\label{fig:MMACAD}
\end{figure}

Details of the development of the Mirror Module Assembly (MMA) are discussed in Ref.~\citeonline{BongiornoSPIE2025}. Testing of the optics is discussed in Ref.~\citeonline{Optics2026}.
A CAD drawing and pictures of the first replicated shell of the MMA are shown in Figure \ref{fig:MMACAD}.  The aluminum mandrels were originally produced for the Micro-X rocket payload but were never polished.  The Marshall team completed polishing last year to a roughness specification of 10 Angstroms.  They have begun shell replication for the first of the five  1 mm thick mirror shells and evaluation of the first shell produced is in progress.  The designs for the support structures for the MMA have been completed and released for fabrication.  The assembly is expected to be completed in early Spring 2027.  Integration of the optics with the rest of the payload will be completed at MSFC in summer 2027.  This  will also afford the team the opportunity to complete an end-to-end test of the payload utilizing the Marshall 100 meter beamline.  We intend to use a single spacing multilayer tuned to 277 eV (the Carbon-K line) installed in the crystal box initially produced for testing the IXPE optics to produce a polarized source for testing REDSoX.  An optical cart and access port have been manufactured and tested for previous sounding rocket missions and will be available for our use.

\vskip -0.1in
\section{Critical-Angle Transmission Gratings}
The CAT gratings produced for REDSoX are 200 nm period, 4 micron deep gratings etched from silicon wafers.  The grating bars are supported by an outer 1 mm thick frame, a hexagonal support structure, and linear supports within those hexagonal cells.  The details of fabrication of these gratings are discussed in Ref.~\citeonline{CATGratingsSPIE_2025}.  Each individual grating is roughly 1 cm by 3 cm and there will be 48 co-aligned gratings on the payload, as shown in Figure \ref{fig:REDSoXEndview}.  

\subsection{Grating Testing}
The initial testing plan for REDSoX gratings involved testing a small subset of the gratings to determine if further testing was statistically necessary.  Testing of initial grating samples showed variation in the efficiencies based on the details of the fabrication procedures in addition to the initial locations of each grating on the larger wafer from which they were etched, and variations that seem to be inherent to the recent history of the individual tools used in various steps of the grating fabrication process.  This led to the realization that unintentionally concentrating particularly efficient (or particularly inefficient) gratings in one petal could lead to biasing of the centroid on the imaging detector. As a result, we made the decision to test each grating individually, so that we could arrange the gratings uniformly between the petals without bias.  

\begin{figure}
    \centering
    \includegraphics[width=13 cm]{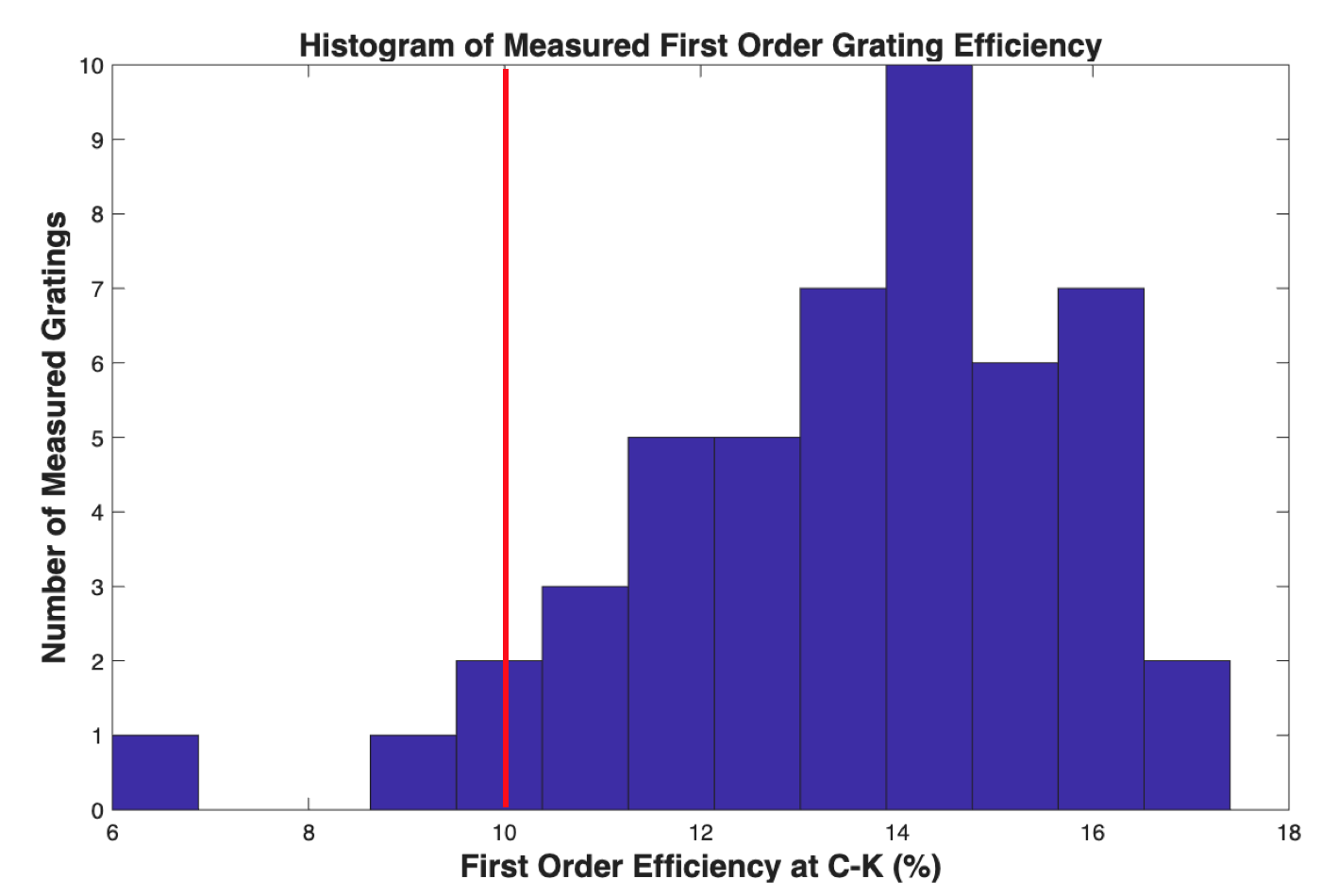}
 \caption{A histogram of the measured average 1st order diffraction efficiencies (all support structure blockage included) for the first 60 gratings evaluated in the beamline.  Measurements are made using the C-K line at 277 eV.  The threshold for acceptance for flight is marked in red.}
\label{fig:eff}
\end{figure}

Our initial testing procedures required roughly a week to test each grating, which is impractical on a mission requiring 48 such gratings.  In order to streamline the process, we developed a testing procedure that can be completed in less than a day for each grating, with only roughly 10-15 minutes of that time being active operator and analysis time.  The beamline testing is automated via custom Labview software, while the analysis is done via Python scripts.  The gratings are tested only in C-K (277 eV) light.  An initial exposure is taken with the grating out of the beam to establish the beam strength.  The grating is moved into the beam and a blaze scan is performed, collecting an image of the first order diffraction at each of several blaze angles of the grating with respect to the incoming light.  From this test, a ``best'' blaze angle is determined.  This test is completed in the center of the grating.  We then hold the blaze angle constant while stepping along the long dimension across the grating to measure the first order diffraction efficiency across the surface of the grating, from which we can produce an average efficiency number for each grating.  We set a rough acceptance criteria of 10$\%$ average efficiency to qualify a grating for flight.  Of the 80 gratings that have been produced and delivered to the REDSoX team, 60 of them have been tested in the beamline and 57 of them have met the acceptance criteria of greater than $10\%$ efficiency.  A histogram of the measured efficiencies of these 60 gratings is shown in Figure \ref{fig:eff}.  We are in the process of testing the remaining gratings, both so that we can select the most appropriate gratings for REDSoX, and also because some of these gratings will be used for the GOSoX mission.  However, we now have sufficient gratings tested and verified to provide for the entire REDSoX grating assembly with spares.  

\subsection{Grating Module Assembly}
\label{sec:gratingmoduleassembly}
Each grating will be epoxied on a titanium mount with a flexure allowing for precise control of the tip (change of blaze angle) of the grating.  The gratings are organized into petals that allow for precise adjustment of the tilt of each grating.  The petals are placed precisely in the converging beam at locations determined by measurements of the slope of the grading of the individual multilayers.  Coalignment of the gratings within the petals will be achieved via a system utilizing a UV laser bounced off of the surface of each grating and reflected into a quadcell detector to determine the angle of that surface.  Details of the assembly and testing of this system in our lab are discussed in \citenum{GratingalignmentSPIE2026}.

We intend to evaluate the alignment of the petals to the rest of the optical elements in the payload by utilizing a portable coordinate measuring machine available at MKI, which can measure relative locations with a precision of microns.

\begin{figure}
    \centering
    \includegraphics[width=15 cm]{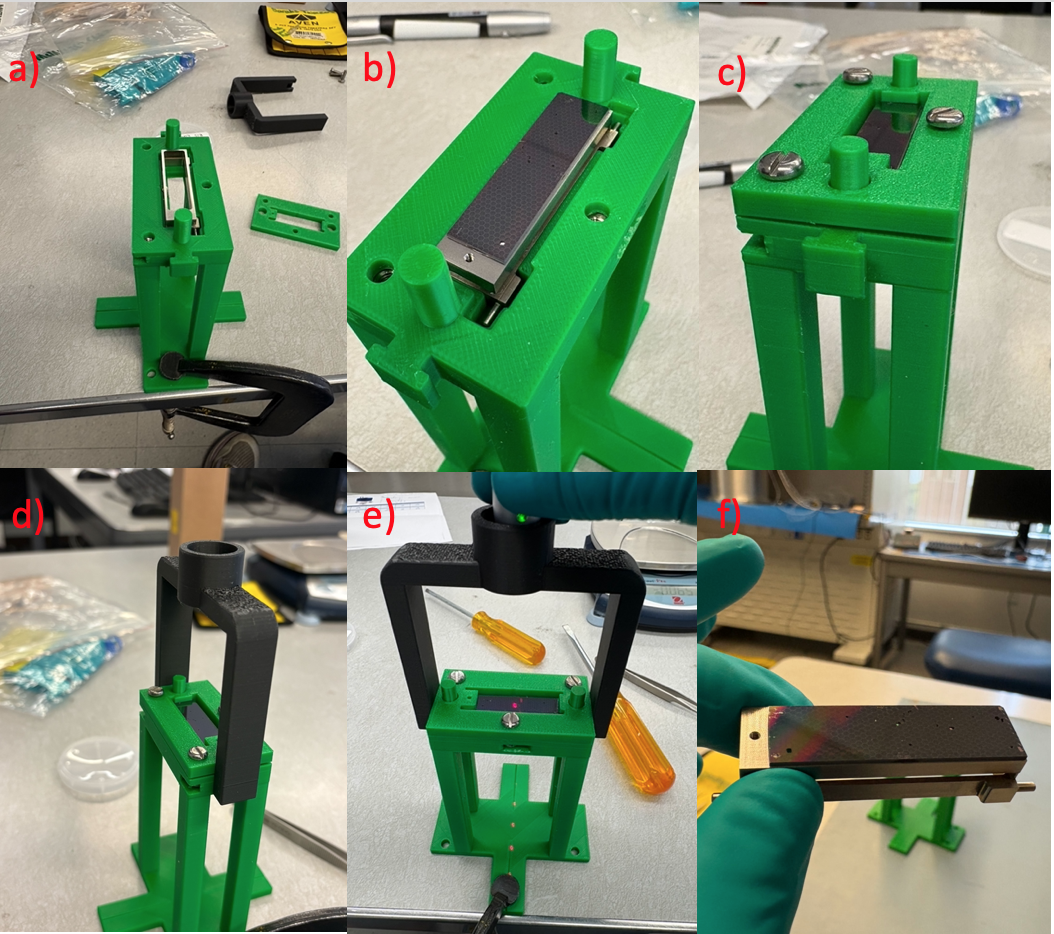}
 \caption{Pictures demonstrating the process of gluing a flight grating to its mount.  a) A mount is placed into the jig, which is a tight press fit.  Several small dots of epoxy are applied to the edges of the frame. b) A grating is gently placed on the mount with its 1 mm silicon frame roughly aligned with the mount frame. c) The clamp is placed on top and bolts are installed without providing any clamping force. d) The laser mount piece is installed.  e) A laser is mounted and shined down through the grating and the grating is gently nudged into alignment with curved tweezers on its outer edges, then the clamp is tightened and the jig is placed in the dry box for the 24 hour cure time of the epoxy.  f) The mount with the attached grating is gently pried out of the jig before being placed in a custom 3D printed storage box.}
\label{fig:jig}
\end{figure}

As a prelude to assembly and alignment of the grating module, we have begun gluing flight gratings to mounts.  We utilize a 3D printed jig that allows us to rigidly attach the grating mount with respect to a line underneath it.  A laser is mounted above the grating and shines through it, producing diffracted laser spots as the visible light diffracts off of the 5-micron period support mesh for the grating bars (the mesh bars are perpendicular to the grating bars).  By aligning those laser spots to the line beneath the grating itself by adjusting the grating rotation with respect to the mount, we can assure the rotation of the grating bars with respect to the mount is within the 1 degree tolerance required for REDSoX.  Images of the current gluing process are shown in Figure \ref{fig:jig}.

We previously mounted and shook a grating to sounding rocket flight levels with X-ray testing before and afterwards to verify that the mounted grating could survive flight \cite{REDSoXSPIE2025}.  In order to verify that the current petal design will survive flight with grating co-alignment intact, we built a `mini-petal' prototype that held two gratings, aligned them, shook the prototype, then checked alignment afterwards showing no change in alignment.  Details of this test are described in Ref.~\citeonline{GratingalignmentSPIE2026}.

\begin{figure}
    \centering
    \includegraphics[width=15 cm]{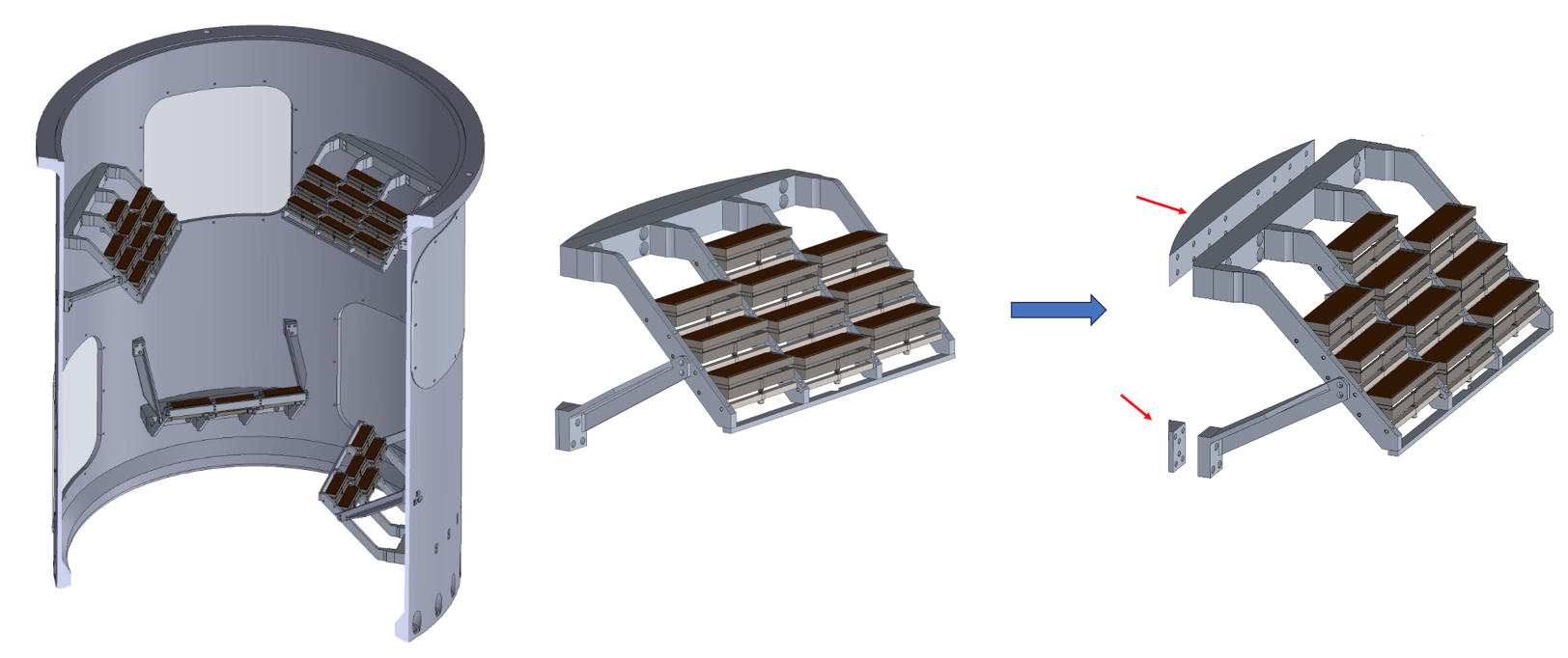}
 \caption{\textit{Left}: A CAD rendering of the grating module with 4 of the 6 total grating petals shown. \textit{Right}: A CAD rendering of an upper grating petal shown with and without its replaceable mounting plate.}
\label{fig:petalmounting}
\end{figure}

The mounting location for each petal will depend precisely on the as-built grading slope of its corresponding multilayer.  While three of the flight multilayers we received had very similar slopes, the fourth (nominally a spare) has a significantly different slope.  We were concerned that, if we needed to swap in the spare, we would need to significantly change the location of the grating petal mounting.  In order to retire this as a risk, the petals have been designed to interface to the cylindrical grating module via mounting plates.  These plates can be remachined and swapped out with new hole patterns to easily change the mounting location of a petal without having to disassemble the petal or re-align the gratings.  A CAD rendering of the grating module as well as a diagram of the mounting between the petals and the module is shown in Figure \ref{fig:petalmounting}.

\vskip -0.1in

\section{Laterally Graded Multilayer Mirrors}
\begin{figure}
    \centering
    \includegraphics[width=15 cm]{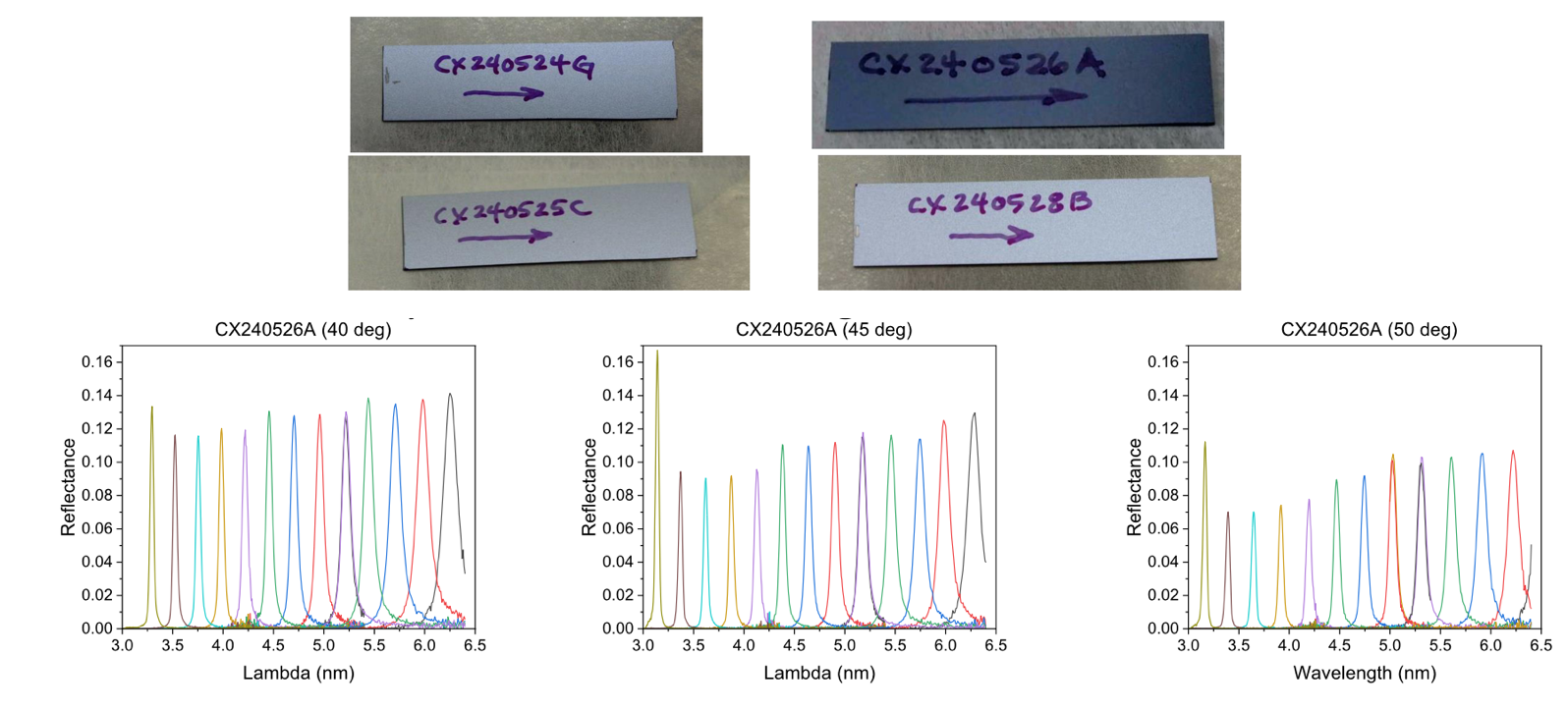}
 \caption{\textit{Top}: A picture of the flight multilayers delivered to MIT from LBNL. \textit{Bottom}: ALS measurements of Bragg peaks on a single multilayer at incidence angles of 40, 45 and 50 degrees.}
\label{fig:asbuiltLGMLs}
\end{figure}

Fabrication of the flight LGMLs at LBNL is complete and they have been tested at the Advanced Light Source (ALS).  The four mirrors (three flight and one spare) have been delivered to MIT along with the data from the ALS, which will allow us to choose the precise location of the gratings with respect to each as-built multilayer in the design, using the formalism previously established \cite{redsoxjatis}.  The mirrors (marked on the back sides) and the data from one of the mirrors taken at the ALS are shown in Figure \ref{fig:asbuiltLGMLs}.  As discussed in Section \ref{sec:gratingmoduleassembly}, the ability to adjust the position of the grating petals has been designed into the payload.  


\begin{figure}
    \centering
    \includegraphics[width=15 cm]{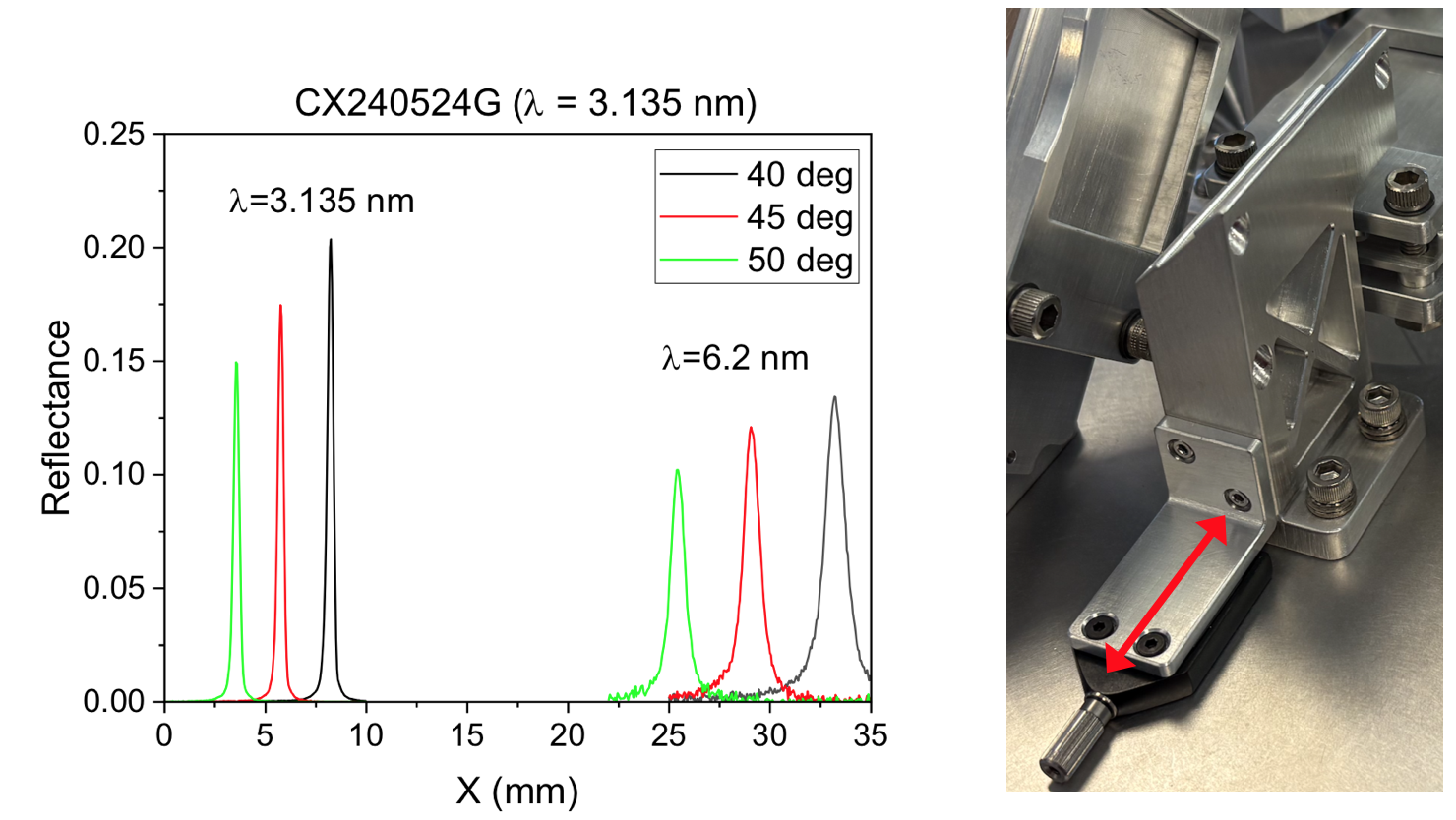}
 \caption{\textit{Left}: The locations of roughly 400 eV Bragg peak at 40 and 50 degree incidence on a single multilayer \textit{Right}: The multilayer adjustment mechanism on a prototype version of the focal plane.  The arrow shows the direction of movement of the mechanism.  Once adjustment is complete bolts will be installed into a slot on the multilayer mount to secure its location for flight.\label{fig:LGMLadjustment}}

\end{figure}

\begin{figure}
    \centering
    \includegraphics[width=6 cm]{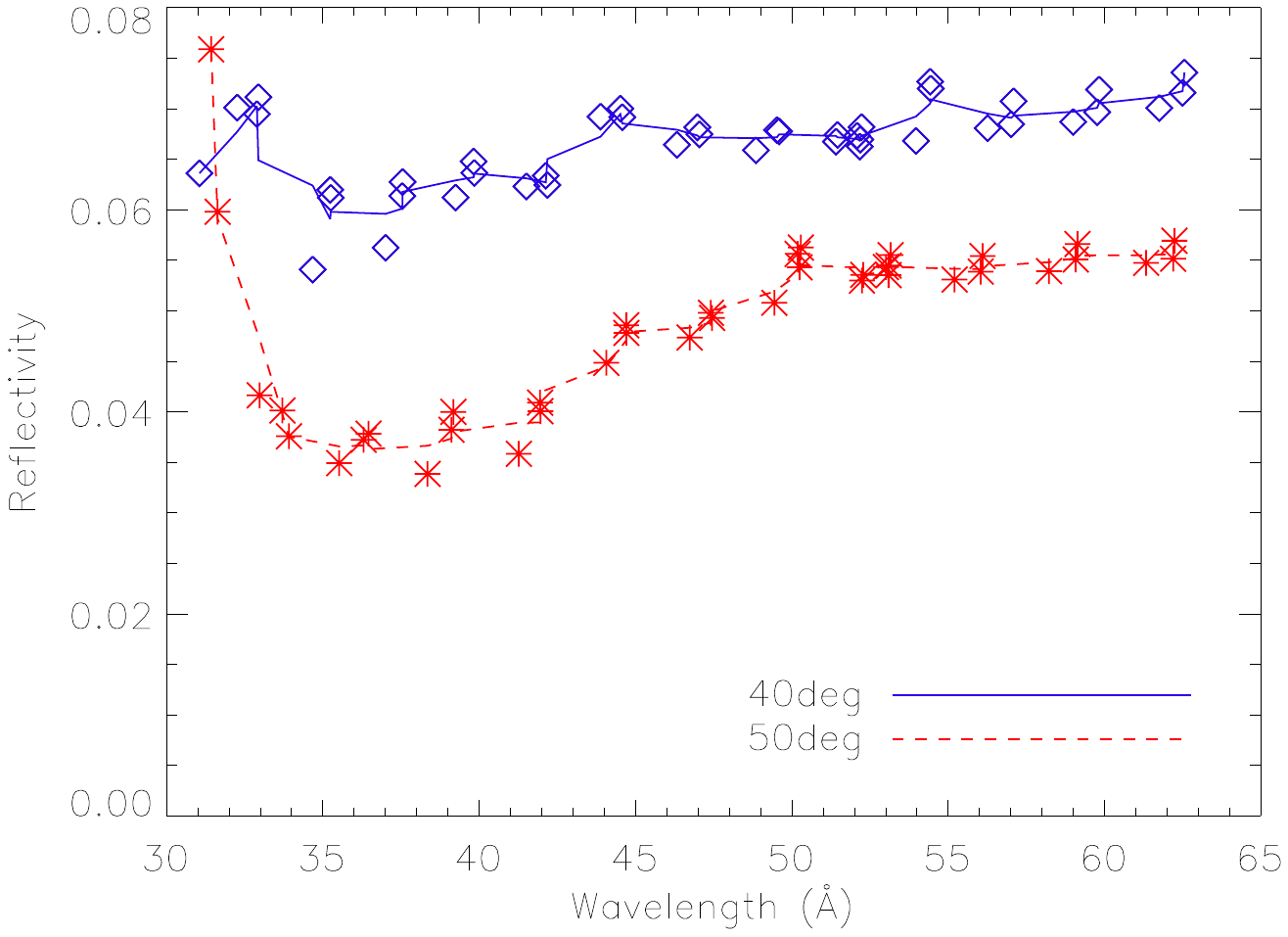}
    \includegraphics[width=6 cm]{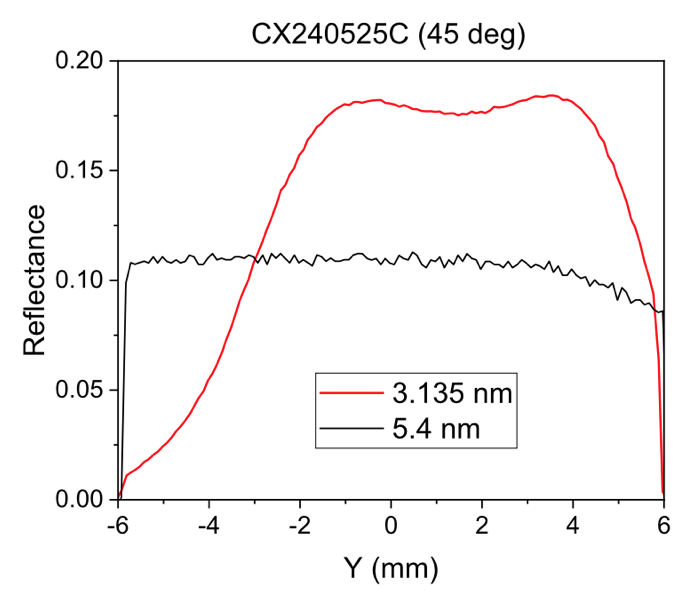}
 \caption{{\it Left:} Reflectivities to unpolarized X-rays at two incident angles, averaged over the four LGMLs.
 {\it Right:} ALS measured reflectance as a function of Y (cross-dispersion) location on the LGML.}
\label{fig:LGMLcrossdispersion}
\end{figure}

We have also designed the focal plane with the ability to move the LGMLs radially by 2-3 mm, which will allow us to adjust their positions should the distance between the 200 eV Bragg peak and the edge of the LGML prove to be different than expected.  The measurements of the locations of the Bragg peaks with respect to the edge of the LGML were measured at the ALS (shown in Figure \ref{fig:LGMLadjustment}); however given the slim tolerance presented by the narrow Bragg peaks we intend to remeasure the locations by installing the multilayers in place of our standard multilayer in the source chamber of the polarimetry beamline, and have also designed the mounting of the multilayers to allow for a slight radial adjustment of 2-3 mm should we discover after assembly that the position of the multilayer is incorrect.  The Bragg peak data from the ALS along with the mechanism for shifting the LGML are shown in Figure \ref{fig:LGMLadjustment}.
Measurements (shown in Figure \ref{fig:LGMLcrossdispersion}) show less than 2-3 \% deviation across 6 mm of the LGML including the center line, which will be sufficient to include the locations to which light will be dispersed.

\section{Detectors}
\begin{figure}
    \centering
    \includegraphics[width=8 cm]{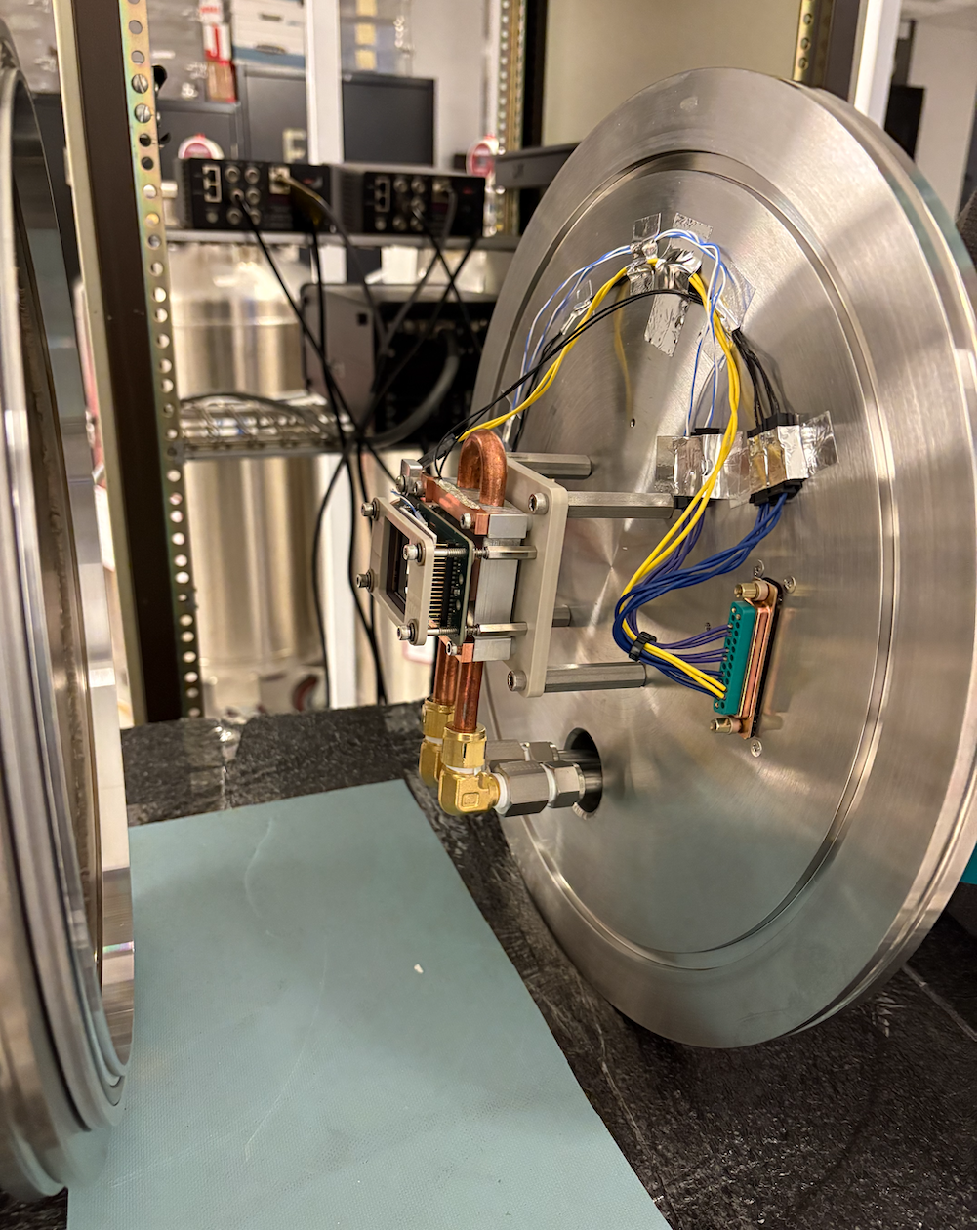}
 \caption{The engineering unit detector installed on its custom testing flange that interfaces with the beamline.}
\label{fig: EMunit}
\end{figure}

A great deal of progress has been made on the functionality of our detectors.  An engineering unit detector and associated readout boards were delivered by XCAM to MIT in December 2024.  We designed and fabricated a test cradle for the detector allowing us to cool it with liquid nitrogen run through a cold block in contact with the back surface of the detector package.  The test cradle is mounted on a vacuum flange that interfaces with the beamline. A ceramic flange electrically isolates the detector flange from the rest of the beamline and its pumps and other instrumentation.  We wrote custom Labview software to control a solenoid valve that turns the liquid nitrogen flow on and off based on the slope of the cooling curve of the detector.  Utilizing this method we can optimize for the fastest cooling while ensuring the cooling slope will stay below 5 K/minute, the rate at which damage may occur.  A picture of the EM unit detector installed on the testing flange in the process of being mated with the beamline is shown in Figure \ref{fig: EMunit}.

We have completed initial testing in this configuration and have written a testing and quicklook tool to allow much faster troubleshooting while running the detector.  We have also acquired our flight computer, a rugged computer from RTD Embedded Technologies Inc., that is designed to withstand sounding rocket vibration levels.  

With the EM unit detector we were able to sample the dark current at various temperatures in order to determine the operating parameters for our detectors in flight.  We found that while dark current increases as expected above -70 C, its level stabilizes below that, so it is not worthwhile to cool below -70 C for flight.  The bulk of our testing was done at -50 C, which is the highest temperature at which we achieved the desired total noise (dark current combined with read noise).  

\begin{figure}
    \centering
    \includegraphics[width=15 cm]{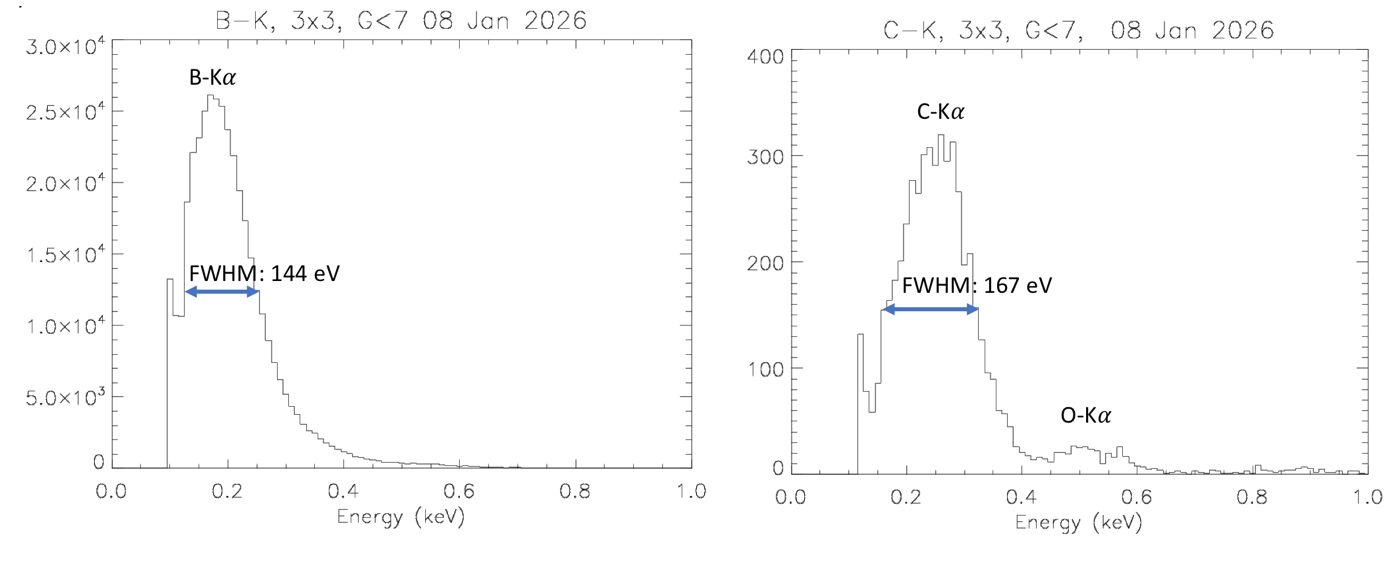}
 \caption{Spectra of the Boron-K line (left) and the Carbon-K line (right) measured by the REDSoX EM unit detector on the MIT polarimetry beamline.  The FWHM values meet the requirement to be less than 200 eV in order to limit background.}
\label{fig: spectra}
\end{figure}

Our principal requirement on the detector is the ability to distinguish the soft X-ray events we utilize for science from the noise floor.  For this we require below 6 electrons of read noise.  After testing various grounding configurations and testing multiple DC-DC converters for distributing power, we were able to run the EM unit with a total noise of roughly 6 electrons.  We installed the EM unit on the end of the beamline and illuminated it with both Boron-K (187 eV) and Carbon-K (277 eV) photons to test that we could distinguish the softest end of our waveband from the noise floor of the detector.  The spectra from both of these tests are shown in Figure \ref{fig: spectra}.  Both the Boron and Carbon lines are easily distinguishable from the noise floor.

We were also able to measure the QE of the EM unit detector relative to that of the reference detector in the beamline.  We measured at a range of energies bracketing the \rs\ bandpass and found it to meet the mission requirements.

We were also able to connect the EM unit detector to the flight computer and read it out using a preliminary version of the flight software, demonstrating our end to end detector readout chain. Due to relatively slow readout of the detectors and a relatively high telemetry rate, we will be able to telemeter full frames to the ground during flight.  Software is in progress to allow us to do event finding and quicklook processing on the ground for the repointing procedure.

XCAM has completed fabrication of the flight units and is currently testing them as a single system with similar grounding and timing to that which will be used in flight.  We expect delivery of the full flight system sometime in mid to late July 2026.

\section{Thermal Design and Prototype Fabrication and Testing}

For the \rs\ flight, we will need to cool the detectors to somewhere between -50 and -70 C on the ground, and keep their temperature within that range during flight.  The flight thermal design calls for cooling the plate on which the detectors and LGMLs are installed with liquid nitrogen to a temperature 5-10 degrees below the desired detector temperature.  This will be achieved with liquid nitrogen flowing through a cold block attached to the plate.  This assembly will also be connected to a buffer plate, which acts solely as extra thermal mass to stabilize the temperature during flight.  The nitrogen flow will be controlled by a solenoid driven by control software similar to that utilized for the EM unit testing that adjusts the solenoid open intervals in response to the temperature and cooling slope of the plate.  Each detector will then have a trim heater installed on its mount, and the individual detector temperatures will be controlled on the ground via a multi-channel Lake Shore thermal controller.

\begin{figure}
    \centering
    \includegraphics[width=15 cm]{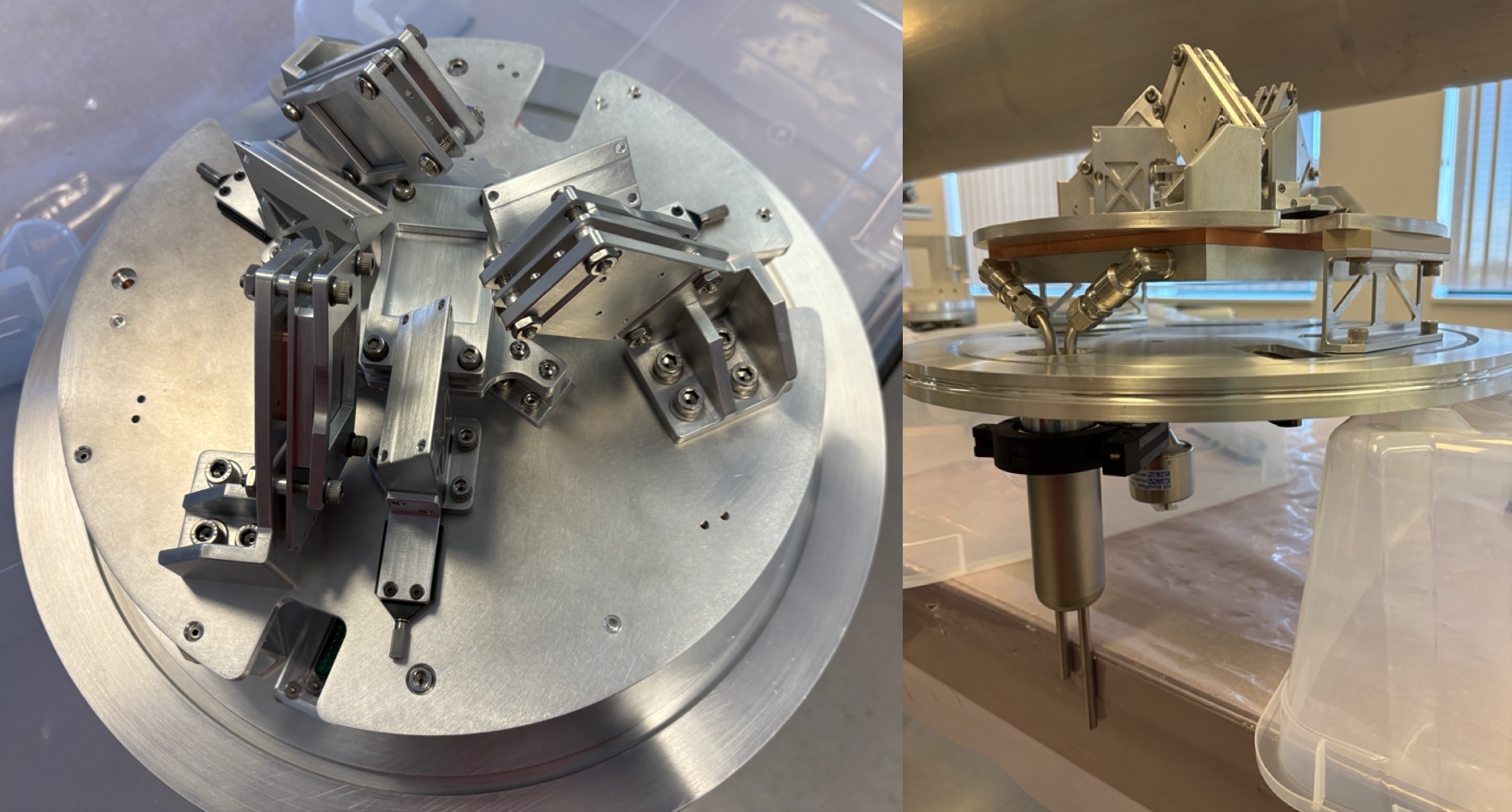}
 \caption{\textit{Left}: A top view of the focal plane prototype. \textit{Right}: A side view of the focal plane prototype installed on a vacuum flange showing the liquid nitrogren vacuum feedthrough lines connected to the cold plate, and the copper buffer plate.  Thermometers and heaters are not installed on the prototype in these images.}
\label{fig: focalplaneproto}
\end{figure}

Upon launch the liquid nitrogen lines and the umbilical line connecting the Lake Shore controllers to the detector trim heaters will pull away and temperature will be maintained solely by the thermal mass of the focal plane plate, cold plate, and buffer plate.  In order to test the thermal design of the focal plane we have constructed a focal plane prototype.  There are aluminum standins for all of the optical elements and cameras, while thermometers and heaters are wired to the detector mounts as they will be in flight.  The plate itself is sized down and a few elements shifted in location to allow the focal plane to fit onto a flange that can be installed on the beamline to allow for in-vacuum testing.  Pictures of the assembled focal plane prototype are shown in Figure \ref{fig: focalplaneproto}.  

We tested the thermal design of the focal plane by installing the prototype on the beamline, cooling the focal plane plate, and controlling the temperature of two of the detectors with two seperate Lake Shore temperature controllers.  We were easily able to control the temperatures of these detector stand-ins to within half a degree showing that our cooling scheme is feasible.  We intend to test the ability of the current design to hold temperature during flight by installing more flight-like wiring (including wiring between the detector stand-ins and the room temperature bulkhead plate), along with heaters on each detector stand in to be able to simulate the heating power created by the running detectors.  Under these conditions we intend to shut off the liquid nitrogen cooling and verify that the detectors can hold temperature sufficiently during a fifteen minute flight-like interval.  This will indicate whether the current buffer plate is sufficient or if we need to add thermal mass to the system to maintain temperature during flight.  

\section{Upcoming work}
Our group continues to work with the NSROC team at WFF to continue to refine and finalize designs and interfaces between the systems.  Cleanroom facilities are being prepared in our lab for assembly of flight parts, and fabrication and assembly of flight parts and assemblies will ramp up in the coming months as final designs are peer reviewed and released.  The payload will be assembled without the optics, then shipped to MSFC for optics integration and end-to-end testing at the 100 meter beamline.  Integration of the optics and end-to-end testing are expected sometime in the summer of 2027, followed immediately by integration at Wallops Flight Facility in the fall of 2027 and flight in late-2027 out of White Sands Missile Range in New Mexico.  The group is also ramping up efforts on the newly-selected GOSoX polarimeter \cite{GOSoXSPIE2026}, which will be strongly based on the final design of REDSoX.  The beginning of the GOSoX program will run concurrently with the final stages of REDSoX development, allowing us to continue to use lessons learned in the REDSoX program to guide the GOSoX design and implementation.
\bibliography{polarimetry20}   
\bibliographystyle{spiebib}   
\include{bibliography}
\end{document}